# Photo-accelerated water dissociation across one-atom-thick electrodes

J. Cai[1,2,3], E. Griffin[1,2], V. Guarochico-Moreira[1,2,4], D. Barry[1], B. Xin[1,2], S. Huang[1,2], A. K. Geim[1,2], F. M. Peeters[5], M. Lozada-Hidalgo[1,2]

[1]National Graphene Institute, The University of Manchester, Manchester M13 9PL, UK
[2]Department of Physics and Astronomy, The University of Manchester, Manchester M13 9PL, UK
[3]College of Advanced Interdisciplinary Studies, National University of Defense Technology, Changsha, Hunan 410073, China
[4]Escuela Superior Politécnica del Litoral, ESPOL, Facultad de Ciencias Naturales y Matemáticas, P.O. Box 09-01-5863, Guayaquil, Ecuador
[5]Departement Fysica, Universiteit Antwerpen, Groenenborgerlaan 171, B-2020 Antwerp, Belgium

Recent experiments demonstrated that interfacial water dissociation ($H_2O \leftrightarrows H^+ + OH^-$) could be accelerated exponentially by an electric field applied to graphene electrodes, a phenomenon related to the Wien effect. Here we report an order-of-magnitude acceleration of the interfacial water dissociation reaction under visible-light illumination. This process is accompanied by spatial separation of protons and hydroxide ions across one-atom-thick graphene and enhanced by strong interfacial electric fields. The found photo-effect is attributed to the combination of graphene's perfect selectivity with respect to protons, which prevents proton-hydroxide recombination, and to proton transport acceleration by the Wien effect, which occurs in synchrony with the water dissociation reaction. Our findings provide fundamental insights into ion dynamics near atomically-thin proton-selective interfaces and suggest that strong interfacial fields can enhance and tune very fast ionic processes, which is of relevance for applications in photo-catalysis and designing reconfigurable materials.

A unique combination of properties in graphene allows using it as a proton-permeable electrode. One-atom-thick graphene exhibits high in-plane electric conductivity[1] and relatively easy proton transport through its basal plane[2-4]. It is also impermeable to all atoms[5-7] and all other ions[8,9] and has exceptional mechanical strength[10]. A recent work reported interfacial water dissociation[11] ($H_2O \leftrightarrows H^+ + OH^-$) through graphene electrodes[12]. These electrodes allow measuring the intrinsic proton currents arising exclusively from the dissociation reaction while experimentally monitoring the interfacial electric field, $E$. The proton currents were found to be exponentially accelerated with increasing $E$ (that reached above $10^8$ V m$^{-1}$), a phenomenon known as the Wien effect. In particular, graphene's perfect selectivity with respect to protons and its atomic-thickness were crucial to observe the Wien effect. These properties enable the intense interfacial $E$ to separate protons from OH$^-$ ions across the atomically-thin barrier that prevents their recombination, thus yielding notable proton currents. The timescale of the involved separation process should be extremely fast – as a first approximation, comparable to the timescale of proton transport and proton-OH$^-$ recombination in water, which is in the sub-picosecond range[13]. On the other hand, previous experiments showed that proton transport through graphene electrodes is strongly enhanced under illumination via a hot-electron mediated mechanism, the so-called photo-proton effect[14].

Hot electrons in graphene have a lifetime of ~1 ps[15,16]. If the proton-hydroxide ion separation across graphene is comparatively fast, then in principle, the photo-proton effect should also accelerate the transport of protons generated by interfacial water dissociation. In this work, we report such acceleration.

Proton-permeable graphene electrode devices were fabricated using monocrystalline graphene obtained by mechanical exfoliation, as reported previously[2,14]. In brief, the crystals were suspended over holes (10 µm in diameter) etched in silicon-nitride substrates[2]. The resulting graphene film was electrically connected to allow for its use as an electrode (Figure 1 and Figure S1). One side of the suspended graphene was decorated with Pt nanoparticles deposited via electron beam evaporation, which served to increase graphene's proton conductivity[2]. The opposite side of the suspended graphene electrode faced a 1M KCl electrolyte with alkaline pH solution (typically, pH 11). The high KCl concentration ensures that the electrolyte resistivity is negligible, whereas the alkaline pH ensured that water is the only source of protons in the system. Hence, in this setup all proton currents arise from the water dissociation reaction, as demonstrated previously[12]. The inner side of the devices was also coated with an anion-exchange polymer (FAA FumaTech) which is an excellent $OH^-$ ion conductor[17]. The polymer coating was not essential for the described experiments; however, it provided additional mechanical support for the membrane, improving the devices' reliability ('Device fabrication' in Supporting Information). For electrical measurements, the devices were connected in an electrical circuit as shown in Figure 1a, using a Pt counter electrode and a silver/silver-chloride reference electrode. All potentials below are referred against the latter electrode, unless stated otherwise. Measurements were carried out inside a chamber with Ar environment and the electrolyte was saturated with Ar to avoid a parasitic oxygen reduction reaction ('Electrical measurements' in Supporting Information).

The rationale to measure interfacial water dissociation ($H_2O \leftrightarrows H^+ + OH^-$) with these devices was demonstrated in ref.[12]. In brief, the dissociation reaction generates protons that transport through graphene. The protons are then adsorbed on Pt nanoparticles by combining with electrons ($H^+ + e^- \rightarrow H^*@Pt$) that flow into graphene through the electrical circuit. These adsorbed protons eventually escape as hydrogen molecules ($2H^* \rightarrow H_2$; Pt catalyses this reaction) through the discontinuous Pt film[2]. Note that the µm-sized electrodes ensure that proton transfer through graphene dominates the resistivity in the circuit, with negligible contributions from the bulk electrolyte and counter electrode[18,19] ('Electrical measurements' in Supporting Information). On the other hand, the photo-effect in proton transport through graphene electrodes was demonstrated in ref.[14]. In that work, conceptually similar devices were measured using an acidic polymer electrolyte, which unlike alkaline electrolytes, contains free bulk protons. Those experiments found that illumination increased the proton transport rate through the graphene electrode via a hot-electron mediated mechanism. In brief, the discontinuous Pt nanoparticle film in the devices yields a spatially inhomogeneous charge doping on graphene that effectively result in a multitude of in-plane p-n junctions[14]. Illuminating such junctions in graphene is known to produce an in-plane hot-electron mediated photo-voltage via the so-called photo-thermoelectric effect[14,16]. In ref.[14] it was shown that in the presence of an out-of-plane source of protons (the acidic polymer electrolyte), this effect strongly accelerates proton transport through graphene's basal plane. In the present work we exploit this effect to accelerate the transfer through graphene of protons generated by the interfacial water dissociation reaction.

The current density vs voltage (*I-V*) response of the devices with alkaline pH electrolyte was measured both in dark conditions and under solar-simulated illumination of 100 mW cm$^{-2}$ intensity (Oriel Sol3A light source). We found that the potential at zero current, $\varphi$, was negative, in agreement with the previous work[12] (Figure S2). For small applied biases *V* around this potential, the *I-V* response was linear, which allowed extraction of the devices' proton conductivity, $G = I/\Delta V$, where $\Delta V = V-\varphi$. Figure 1 shows a typical *I-V* response of the devices measured at pH 11 under illumination. Surprisingly, *G* increased by an order of magnitude with respect to the dark case. The inset of Figure 1 shows that this photo-response was stable and displayed no signs of deterioration after several hours of continuous illumination. To explore these observations further, we measured the devices using solutions with different alkaline pH (Figure S3). The absolute value of *G* in dark conditions changed with pH, in agreement with the previous report[12]. In all cases, we have observed strong increase in *G* under illumination.

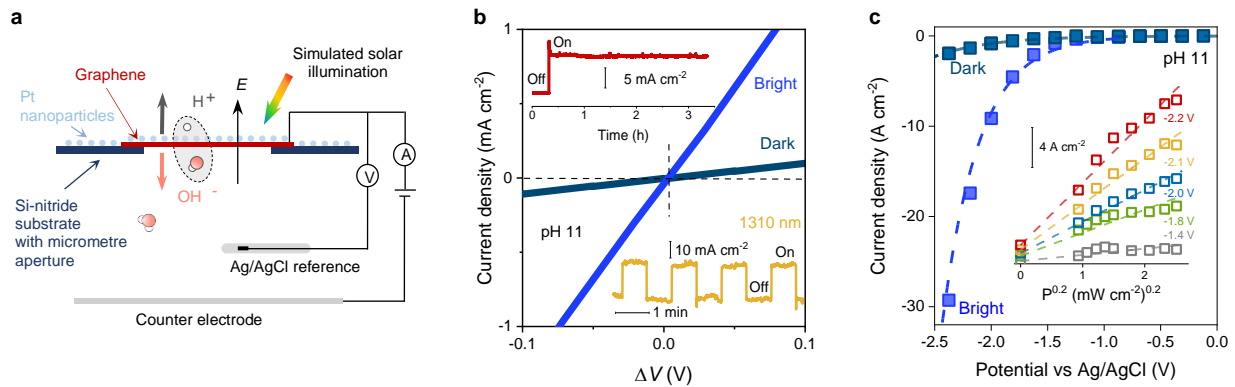

**Figure 1.** Photo-effect in water dissociation at graphene electrodes. (a) Schematic of graphene devices and the measurement setup. Water molecules dissociate under the strong interfacial *E*. Protons transfer through graphene and adsorb on its Pt-decorated external surface, whereas OH$^-$ ions drift into the bulk electrolyte. Red and white balls represent oxygen and hydrogen atoms. (b) Typical *I*-$\Delta V$ characteristics for small $\Delta V = V-\varphi$ around the potential of zero current, $\varphi$ (Figure S2). Response under the dark conditions (dark blue) and solar-simulated illumination of 100 mW cm$^{-2}$ (bright blue). Dashed lines, guides to the eye. The top inset shows that the photo-response was stable for hours of continuous illumination ($\Delta V = 0.2$ V). Bottom inset, current density vs time for illumination using on-off pulses (1310 nm light source with 17 mW cm$^{-2}$ intensity; $\Delta V = 0.4$ V). (c), Examples of *I-V* characteristics away from the linear regime for our devices in dark (dark blue) and bright conditions (bright blue). Dashed lines, guides to the eye. Inset, $I(P) \propto P^{0.2}$ current vs illumination power relation found for devices. Dashed lines, guides to the eye.

The photo-effect was also observed at high biases, away from the linear regime. In those measurements, we fixed the potential of the graphene electrode vs the reference, measured *I* as a function of time and then illuminated the devices in 1 minute long on-off pulses. Figure 1b (top inset) shows that in this high-*V* regime the devices also displayed a strong enhancement of *I* under illumination. To characterise the effect, we measured the dependence of the photo-response as a function of the illumination power density *P* ranging from 0.7 to 100 mW cm$^{-2}$. Figure 1c shows that the found *I(P)* dependence could be described by the empirical relation $I \propto P^{0.2}$, which is consistent with the dependence found for the photo-proton effect reported in ref.[14].

To rule out any possible artefact, we performed additional measurements. Neither the polymer support nor Pt nanoparticles could yield the photo-response reported here (Fig. S6). Besides graphene, this leaves only the silicon/silicon-nitride substrate as an alternative photo-absorber. It is unlikely that the robust photo-effect we observed is due to silicon, since it is insulated from the electrical contacts by a thick 500 nm nitride layer (Figure S1). Moreover, we previously demonstrated that the photo-effect in these devices can be entirely suppressed if metals like Au or Ag were used instead of Pt nanoparticles[14], which would be inconsistent with silicon as the photo-absorber. Nevertheless, we characterised the photo-effect using 1310 nm light source. Such a long wavelength cannot be absorbed by the ~1.2 eV bandgap in silicon, but is readily absorbed by graphene[20]. We also studied these devices using an acidic pH polymer (Nafion) to study the role of this long wavelength on proton transport. Figure 1b (bottom inset) shows that the devices displayed the same strong photo-response even with this long-wavelength light. Devices measured under alkaline pH conditions displayed the same enhancement. These observations prove that the photo-effect is a feature of graphene and its proton transport.

Our results can be understood as follows. The intense electric field at the graphene-water interface, which is of the order[12] of $10^8$ V m$^{-1}$, dissociates water molecules into protons and hydroxide ions ('Wien effect' in Supplementary Information). The same field drives protons through graphene and hydroxide ions into the electrolyte bulk, separating the generated ion pairs across the proton-selective interface that prevents their recombination. The transported proton then adsorbs on a Pt nanoparticles on the opposite side of graphene by acquiring an electron. The role of illumination can be understood using the following two observations. First, while the absolute value of $I$ depends on pH, the photo-effect always enhances $I$ by a factor of ~10 with respect to the dark case. This is the same enhancement reported in ref.[14] for devices in acidic pH in which protons are free in the bulk electrolyte. Second, the illumination power dependence, $I(P)$, reported here is the same as in ref.[14]. These observations are consistent with photo-acceleration of proton transport events via the previously reported photo-proton effect. This is possible for this reaction in graphene electrodes because the intense interfacial electric field acts over atomic scale distances and thus achieves fast separation of the generated protons across graphene – within a timescale comparable to or shorter that the ps lifetime of hot electrons in graphene ('Timescales' in Supplementary Information).

Water dissociation ($H_2O \leftrightarrows H^+ + OH^-$) eventually leads to full electrolysis ($H_2O \rightarrow H_2 + \frac{1}{2}O_2$), producing hydrogen and oxygen gas. The gas evolution rates are much slower than the dissociation step[21] ('Timescales' in Supplementary Information). However, in our devices these reactions take place in the large Pt nanoparticle film ($H_2$ evolution) and the Pt counter-electrode ($O_2$ evolution). These catalytically active areas are several orders of magnitude larger in size than the graphene electrode and hence effectively behave as drain reservoirs for the ions. Because of this, gas evolution reactions are not a limiting factor in our devices[14] and, therefore, we also expect to observe an acceleration of these reactions under illumination. To confirm this, we measured rates of $H_2$ and $O_2$ production directly, both in dark conditions and under illumination. For hydrogen measurements, the graphene electrode faced a vacuum chamber connected to a mass spectrometer, whereas an oxygen-concentration sensor (Clark microelectrode) placed inside the electrolyte solution monitored oxygen production (Figure S4 & S5). For zero or positive voltages applied to graphene, no $H_2$ could be detected by the spectrometer, in agreement with the previous work[6,14]. For the negative polarity, both $H_2$ flux and electric current were detected simultaneously. Figure 2 shows that for every two electrons that flowed through the electrical circuit, one $H_2$ molecule was detected by the spectrometer. This charge-to-mass conservation is described by Faraday's law of electrolysis: $\Phi_{H2} = I/2F$, with $\Phi_{H2}$ the hydrogen flux and $F$ the Faraday constant. For $O_2$ gas, the area-normalised derivative of the oxygen concentration versus time, $d(O_2)/dt = \Phi_O$ was also described by Faraday's law $\Phi_O = I/4F$ (Figure 2). Illuminating the devices resulted in an instantaneous increase in both electrical current and gas flux (Figure S4 & S5) that was also consistent with Faraday's law of electrolysis. The found relations show that $H_2$ and $O_2$ molecules were generated in a 2:1 ratio with 100% Faradaic efficiency, which again shows that the measured currents in our devices are due to water dissociation.

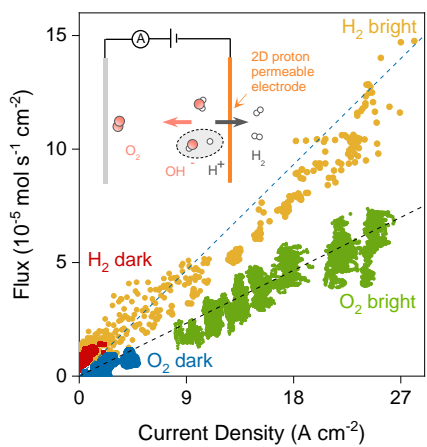

**Figure 2.** Faradaic efficiency measurements. Hydrogen and oxygen fluxes as a function of $I$ under dark and bright conditions; colour coded. The dotted lines correspond to $\Phi_{H2} = I/2F$ and $\Phi_O = I/4F$. Inset, schematic of gas production measurements.

Our work reports a strong photo-effect in the interfacial water dissociation reaction using one-atom-thick graphene electrodes. The observation is attributed to acceleration of proton transport which happens in synchrony with the dissociation reaction. The findings are consistent with both the fast rate of proton transport and proton-$OH^-$ recombination in water (sub-picosecond scale) and the lifetime of hot electrons in graphene (picosecond timescale). We have also shown that strong electric fields acting across atomically thin and highly selective interfaces can enable ultrafast ion-charge separation. The fundamental insights gained here could be of interest for development of photocatalysts, for which charge separation is a central consideration. Another emerging possibility is the use of protons to reversibly modify the electronic properties of materials. This has been explored in low power memory storage[22], plasmonic materials[23] and neuromorphic hardware[24]. The protonation dynamics in these applications is important as it can control the response time in write/read or potentiation/de-potentiation cycles[22-24]. In addition, our work suggests that atomically-thin interfaces can control ion-charge separation dynamics at timescales comparable to those in optoelectronics, which could open new opportunities in these applications.

# Supporting Information for

# Photo-accelerated water dissociation across one-atom-thick electrodes

J. Cai[1,2,3], E. Griffin[1,2], V. Guarochico-Moreira[1,2,4], D. Barry[1], B. Xin[1,2], S. Huang[1,2], A. K. Geim[1,2], F. M. Peeters[5], M. Lozada-Hidalgo[1,2]

[1]National Graphene Institute, The University of Manchester, Manchester M13 9PL, UK
[2]Department of Physics and Astronomy, The University of Manchester, Manchester M13 9PL, UK
[3]College of Advanced Interdisciplinary Studies, National University of Defense Technology, Changsha, Hunan 410073, China
[4]Escuela Superior Politécnica del Litoral, ESPOL, Facultad de Ciencias Naturales y Matemáticas, P.O. Box 09-01-5863, Guayaquil, Ecuador
[5]Departement Fysica, Universiteit Antwerpen, Groenenborgerlaan 171, B-2020 Antwerp, Belgium

**Experimental Section**

**Device fabrication.** Apertures 10 μm in diameter were etched into silicon nitride substrates (500 nm $SiN_x$ on B-doped Si, purchased from Inseto Ltd.), following the protocol previously reported[1]. Au electrodes were fabricated onto the substrates using photo-lithography and electron-beam evaporation. Mechanically exfoliated (monocrystalline) graphene[2] was then suspended over the apertures and on the Au electrodes (Figure S1). Pt nanoparticles were deposited by electron beam evaporation on the graphene film[3]. The discontinuous Pt film had a nominal thickness of 1 nm and is not electron conductive, which allowed using graphene as the electrode. Ranging the film thickness between 0.3 nm and 2 nm had no noticeable impact on the device performance. Thicker Pt films became electrically conductive, which shorted the graphene electrode. Thinner films displayed low proton currents. On the opposite side of the graphene film, an anion-exchange ionomer solution (Fumion FAA-3-SOLUT-10, FuMA-Tech purchased from Ion Power GmbH) was drop cast. The polymer acts as a versatile anion conductor in which both the type of charge carriers and their concentration can be controlled by placing it in contact with an electrolyte solution[4]. The polymer was placed in contact with an Ar-saturated alkaline electrolyte consisting of 1M KCl with KOH to set the pH of the electrolyte (typically 11). All measurements were carried out in a chamber with Ar environment and the electrolyte was saturated with Ar. Before measurements, the polymer side of the devices was left in contact with deionised water for several hours to remove impurities and was then exposed to the electrolyte solution.

The polymer and the Pt nanoparticles are not expected to display a photo-response, however, we corroborated this directly (Fig. S6). To that end, silicon nitride substrates with 10 μm diameter through holes were fabricated as discussed above. Fumion polymer was drop cast on one side. The opposite side was then decorated with Pt nanoparticles deposited by electron beam evaporation and then coated with Fumion (Fig. S6a). The devices were measured in dark conditions and solar simulated illumination. No photo-response was observed, as expected (Fig. S6b).

**Electrical measurements.** For electrical measurements the graphene electrodes were connected in a three-electrode geometry using a Pt counter electrode and an Ag/AgCl reference electrode. A dual channel Keithley SourceMeter 2636A that was programmed to function as a potentiostat was used to measure the potential of the graphene electrode vs the Ag/AgCl (voltmeter channel) and the current between graphene and the Pt electrode (source channel). During the measurement, a feedback unit (a Proportion Integration control loop) between the voltmeter and source channels set the potential vs Ag/AgCl reference into a required setpoint. For reference, we also performed measurements using an Ivium CompactStat.h potentiostat, which gave the same results. We typically scanned $\pm 150$ mV around the zero current potential (vs the reference electrode) at sweep rates of 0.01 V min$^{-1}$. The photo-response of the devices in this work was characterised using a calibrated Newport Oriel Sol3A light source that produced solar-simulated illumination with maximum intensity of 100 mW cm$^{-2}$. The illumination intensity was controlled with the solar simulator's aperture diaphragm.

An important consequence of the small size of the graphene electrode is that the bulk electrolyte resistance does not limit the current measured[5,6]. For a cell with a microelectrode of radius $r$ and electrolyte conductivity $\kappa$, the limiting resistance ($R$) is given by $R = (4\pi\kappa r)^{-1}$. In our measurements, $\kappa \approx$ 0.1 S cm$^{-1}$, $r = 5$ μm and hence $R \approx 1.6$ kΩ, which is at least 2 orders of magnitude smaller than any of those reported with graphene electrodes.

**Faradaic efficiency measurements.** Figure S4 shows a schematic of the experimental setup for Faradaic efficiency measurements, demonstrated in refs.[3,7]. The graphene device separated two chambers. The Pt-decorated side facing the inside of a chamber evacuated and connected to the mass spectrometer. The opposite side of the device faced the electrolyte solution. The hydrogen flux and electric current were measured simultaneously using a mass spectrometer (Inficon UL200 Detector) and a Keithley 2636A sourcemeter. We did this measurement in two ways. First, we applied a fixed voltage and illuminated the device in on-off cycles (Figure S4b). Second, the illumination was turned on and the voltage was swept (Figure S4c). Both methods yielded the same dependence of hydrogen flux versus current density, $\Phi_{H2} = I/2F$, with $F$ the Faraday constant.

Figure S5a shows a schematic of our oxygen flux setup, which was demonstrated in ref.[7]. In brief, a graphene device was clamped to a transparent acrylic container with the polymer side of the device facing the inside of the container. The container had three gasket-sealed outlets for a Clark oxygen microelectrode (UNISENSE, OX-NP), a needle connected with an Argon supply and a Pt wire electrode. A small magnetic stir bar kept at a rotation rate of 300 rpm promoted gas convection in the electrolyte solution. The solution was purged with argon gas through the needle for at least 30 mins and the whole container was placed inside a chamber with constant argon gas circulation to prevent oxygen leakage into the cell. Electrical current and oxygen concentration [$O_2$] in the solution were measured simultaneously. In a typical measurement, voltage is applied to the device, illumination is turned on and then both illumination and voltage are turned off. Figure S5c shows d[$O_2$]/d$t$ from a typical measurement. The area-normalised oxygen concentration $\Phi_O = (d[O_2]/dt)/A$ was correlated with the measured current via the Faradaic relation as $\Phi_O = I/4F$.

The functioning of the oxygen sensor was described in detail in ref.[7]. In brief, the sensor consists of a pipette containing an oxygen-reducing cathode, a reference electrode and a guard electrode. The tip of the sensor is sealed with a silicone membrane, which is impermeable to all ion but highly permeate to gases[8]. This creates a chamber with a stable environment for the electrolyte (Figure S5b). During operation, the potential of the sensing cathode is polarized against the reference electrode and the diffusion of oxygen through the membrane is detected by the sensing cathode, via the oxygen-reducing reaction: $O_2 + 2H_2O + 4e^- \rightarrow 4OH^-$. The resulting pA-level current signal is amplified to convert it to voltage in the mV range. Since the sensing cathode only consumes a negligible amount of oxygen, the guard cathode removes the excess oxygen in the electrolyte.

**Supplementary discussion**

**Wien effect.** In a recent work, we reported that the proton currents arising from interfacial water dissociation through graphene electrodes are exponentially accelerated with increasing $E$[7]. However, we note that graphene's proton conductivity and the Wien effect are independent phenomena. Strong electric fields accelerate the water dissociation reaction, yielding additional protons and hence higher proton currents – a process modelled by Onsager's theory of the Wien effect. In ref. 7, we showed that this generation of additional protons fully explains the field dependence of the water dissociation in graphene electrodes and, hence, that the field effect is not due to an increase in graphene's proton conductivity in strong $E$. Moreover, we note that the proton conductivity of graphene is not expected to depend on the electric field[1]. This is because proton transport is determined by the total density of electron clouds in graphene[1]; whereas, even at the highest doping density used in our experiments, the number of electrons induced by $E$ present only a small portion (<1%) of the total number of electrons in un-doped graphene[7].

**Timescales**. Our experiments provide insights into the timescales of the water dissociation process through graphene as follows. In our devices, illumination accelerates proton transfer via the photo-proton effect[3]. This process relies on the ability of graphene to absorb photons to excite electrons above the Fermi energy – the so-called hot electrons. The timescale during which the electrons remain excited, or hot, is in the pico-second timescale[3]. On the other hand, proton currents in our devices arise from water dissociation, which involves the separation of protons from hydroxide ions across graphene. Hence, the observation of photo-accelerated water dissociation in our devices suggests that the time it takes to separate the proton-hydroxide ion pairs through graphene is comparable to or faster than the lifetime of the hot electrons – otherwise the two processes would be decoupled and no photo-effect would be measured. On this basis, we suggest that the proton-hydroxide separation process takes place in the pico- or sub-picosecond timescale. For reference, we note that, as an independent approximation, the proton-hydroxide ion separation process can be expected to take place within timescales comparable to those of proton transport and proton-OH⁻ recombination in water (sub-picosecond timescale, ref.[9]). This independent approximation is consistent with the timescale estimated from the observation of the photo-effect.

The fast water dissociation ($H_2O \leftrightarrows H^+ + OH^-$) process eventually leads to full electrolysis ($H_2O \rightarrow H_2 + \frac{1}{2}O_2$), producing hydrogen and oxygen gas. The gas evolution rates are slower than the dissociation rate and are generally expected to take much longer than the ps lifetime of hot electrons in graphene[3]. If these reactions were limiting in our devices, we would not observe a photo-response. However, in our devices these reactions take place in the large Pt nanoparticle film ($H_2$ evolution) and the Pt counter-electrode ($O_2$ evolution), which are several orders of magnitude larger than the graphene electrode. In previous work we showed that because of this large size difference, the H$_2$ evolution rate is no longer limiting and the Pt film effectively behaves as drain reservoirs for protons[3]. The observation of a strong photo-response in the water electrolysis reaction in the present work shows that the same holds for the large Pt counter-electrode for O$_2$, as expected.

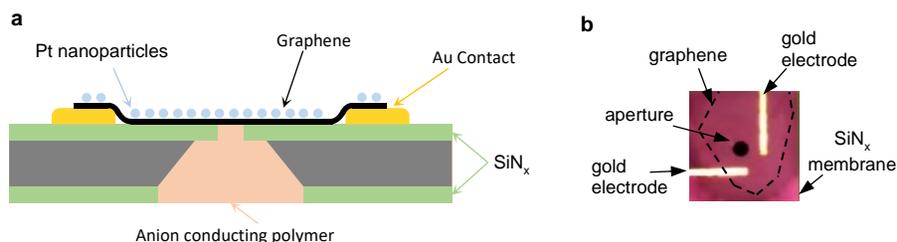

**Figure S1. Device geometry.** (a) Schematic of graphene electrode devices. (b) Optical image of one of these devices (plan view). Black circle, 10 µm diameter aperture in the SiN$_x$ substrate. Dashed lines mark the area covered by monolayer graphene. Devices typically had two gold electrodes (usually they were shorted) in order to ensure good electrical contact with graphene.

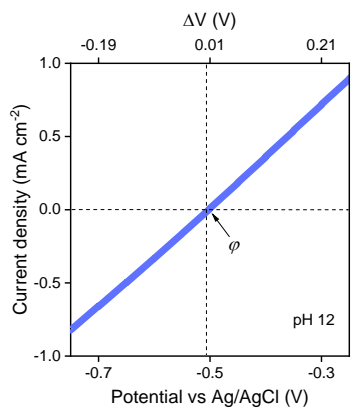

**Figure S2. Current voltage characteristics.** *I-V* response of devices vs reference electrodes. The potential at zero current, $\varphi$, is typically negative, in agreement with previous work[7]. The I-V response is linear for small $\Delta V = V-\varphi$ (top x-axis). Dashed lines, guide to the eye.

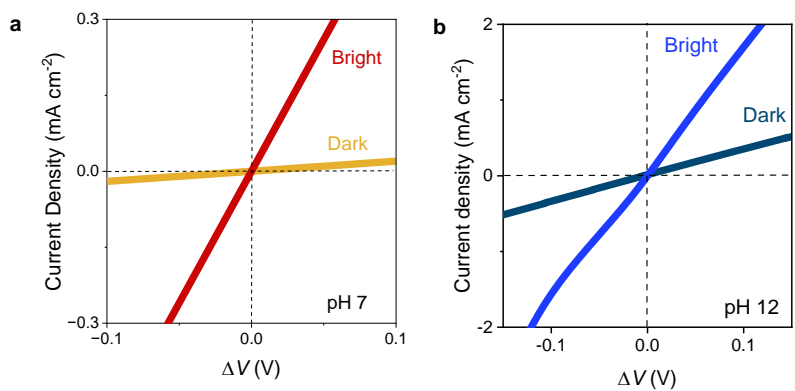

**Figure S3. Photo-effect measured using different electrolyte pH.** Examples of *I-V* characteristics of graphene electrode devices measured in dark and bright conditions using electrolyte pH 7, (a) and pH 12, (b) Solar simulated illumination of 100 mW cm$^{-2}$. The absolute value of the current depends strongly on pH, as reported in ref.[7], but the bright current increases by an order of magnitude with respect to the dark in all cases.

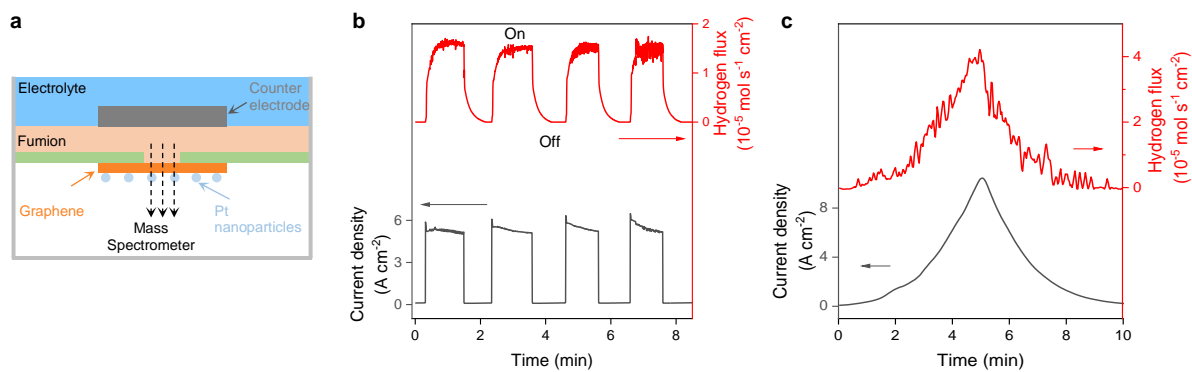

**Figure S4. Hydrogen mass transport experiment.** (a) Schematic of the experimental setup. (b) Example of current density and hydrogen flux measurements recorded simultaneously while switching the illumination on and off. *V*-bias, 1.8 V. (c), Example of current density and hydrogen flux measurements recorded simultaneously with illumination turned on, while sweeping bias voltage (0-2 V). pH 11.

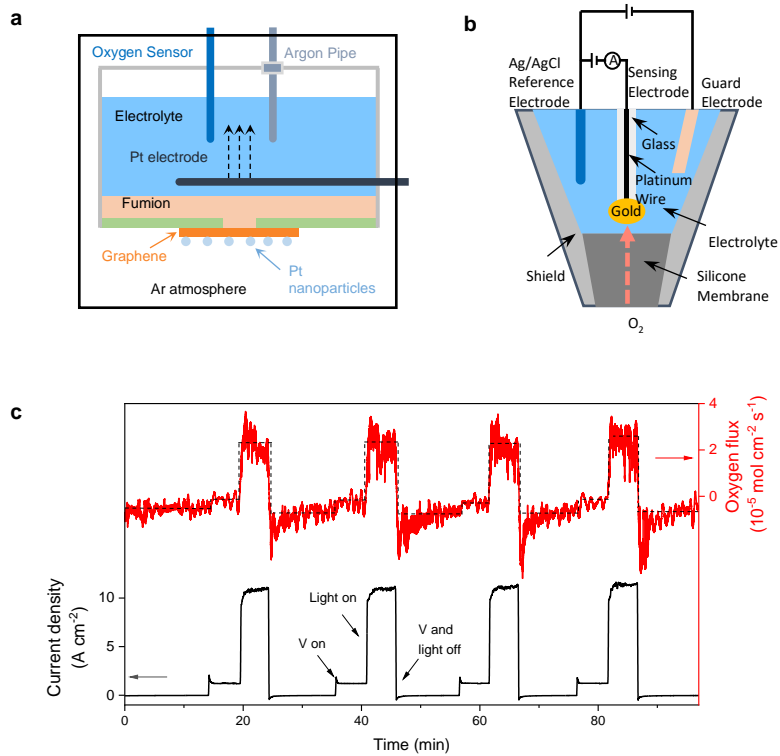

**Figure S5. Oxygen mass transport experiment.** (a) Schematic of the experimental setup. (b) Schematic of the oxygen sensor. (c) Example of current density and oxygen flux data recorded simultaneously while switching illumination on and off, pH 11.   *V*-bias, 2.1 V. Measuring sequence includes four steps: zero bias in dark; voltage applied in dark; illumination switched on; both voltage and illumination switched off. Dotted lines, guide to the eye.

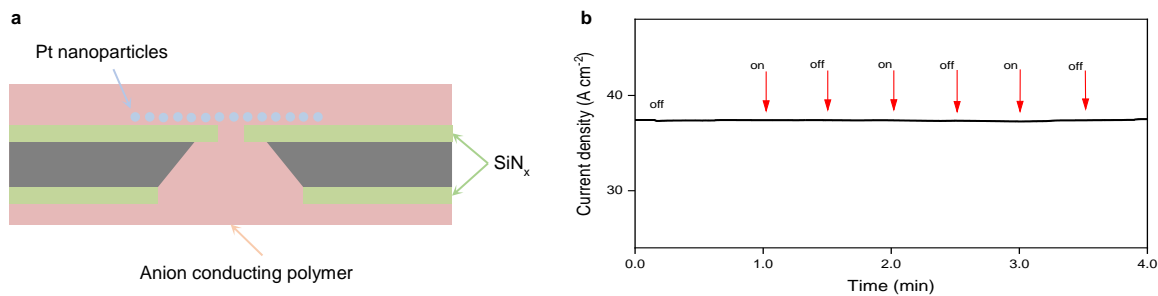

**Figure S6. Absence of photo-response in devices without graphene.** (a) Devices without graphene used to test the photo-response of the polymer and Pt nanoparticles. (b), Current density vs time for a device without graphene electrode in dark conditions and under solar-simulated illumination. The light was turned on and off in 30-second intervals (marked with red arrows). Voltage bias, 0.2 V. No photo-response was observed.